\documentclass[12pt]{iopart} 
\usepackage{graphicx}
\usepackage{bm}
\usepackage{amssymb,amsfonts,amsmath}

\begin{document}
% some user-defined commands ============================
\renewcommand{\vec}[1]{\boldsymbol{#1}}
\newcommand{\GSA}{ground state atom}
\newcommand{\RE}{Rydberg electron}
\newcommand{\density}{\mathcal{N}}
\newcommand{\rr}{\vec{r}}
\newcommand{\RR}{\vec{R}}
\newcommand{\abs}[1]{\left|#1\right|}
\newcommand{\me}{m_\mathrm{e}}
\newcommand{\mRb}{m_\mathrm{Rb}}
\newcommand{\ee}{\mathrm{e}}
\newcommand{\dd}{\mathrm{d}}
\newcommand{\ii}{\mathrm{i}}
\renewcommand{\vec}[1]{\bm{#1}}
\newcommand{\vRb}{\vec{v}_\mathrm{Rb}}
\newcommand{\ve}{\vec{v}_\mathrm{e}}
\newcommand{\AZ}[1]{``#1''}
\newcommand{\psiryd}{\psi_\mathrm{Ry}}
\newcommand{\psipw}{\psi_\mathrm{pw}}
\newcommand{\order}[1]{\mathcal{O}\left(#1\right)}
\newcommand{\momentum}[1]{\vec{p}^{(#1)}}
\newcommand{\vbar}{\bar{v}}
\hyphenation{Ryd-berg}

% Headlines =================================================================
\title{Quantum-classical lifetimes of Rydberg molecules}

\author{Andrej Junginger, J\"org Main, and G\"unter Wunner}

\address{1. Institut f\"{u}r Theoretische Physik, Universit\"{a}t Stuttgart,
70550 Stuttgart, Germany}

\date{\today}

% Abstract =================================================================
\begin{abstract}
A remarkable property of Rydberg atoms is the possibility to create molecules
formed by one highly excited atom and another atom in the ground state. The
first realisation of such a Rydberg molecule has opened an active field of
physical investigations, and showed that its basic properties can be described
within a simple model regarding the ground state atom as a small perturber that
is bound by a low-energy scattering process with the Rydberg electron [C. H.
Greene \etal, Phys.\ Rev.\ Lett.,\ 85:\ 2458 (2000)]. Besides the good agreement
between theory and the experiment concerning the vibrational states of the
molecule, the experimental observations yield the astonishing feature that the
lifetime of the molecule is clearly reduced as compared to the bare Rydberg atom
[B.\ Butscher \etal, J.\ Phys.\ B:\ At.\ Mol.\ Opt.\ Phys.\ 44, 184004 (2011)].
With focus on this yet unexplained observation, we investigate in this paper the
vibrational ground state of the molecule in a quantum-classical framework. We
show that the Rydberg wave function is continuously detuned by the presence of
the moving ground state atom and that the timescale on which the detuning
significantly exceeds the natural linewidth is in good agreement with the
observed reduced lifetimes of the Rydberg molecule.
\end{abstract}

% Pacs ==================================================

% Title =================================================
\maketitle

\section{Introduction}

The field of Rydberg atoms which possess a highly excited valence electron has
been established decades ago, and it is still an active one today (see Ref.\
\cite{Loew2012} for a recent review and references therein). One reason for the
interest in such highly excited atoms is the fact that important properties of
these atoms in general scale universally with powers of the principal quantum
number $n$. For highly excited atoms ($n\gg 1$) this leads, e.g., to extremely
large extensions up to the size of a virus, huge polarizabilities as well as van
der Waals-coefficients and long lifetimes, making these atoms ideal candidates
for applications in quantum simulations or quantum computing \cite{Saffman2010}.

Besides these remarkable properties, Rydberg atoms are also of great interest,
because they are able to form very weakly bound molecules together with a second
atom in the ground state \cite{Greene2000}. Such Rydberg molecules were first
realised in 2009 with rubidium atoms \cite{Bendkowsky2009}, and the experiment
showed that the basic properties of the molecule can be well described within a
simple model \cite{Greene2000}. Regarding the \GSA\ as a small perturber which
is polarised by the Rydberg electron the interaction can be described
theoretically by a low-energy scattering process. If the respective scattering
length is negative, the interaction leads, within a mean-field approximation, to
a binding and oscillatory molecular potential, and the theoretical predictions
agree very well with the measured vibrational spectra of the molecules
\cite{Bendkowsky2009}.

However, a remarkable and not yet fully explained property of the Rydberg
molecules is that surprisingly they exhibit mean lifetimes $\tau$ which are
clearly reduced when compared to the bare atomic Rydberg state
\cite{Bendkowsky2009, Greene2009, Browaeys2010}. Butscher \etal\
\cite{Butscher2011} found that the atomic background gas has a significant
influence on the lifetime due to inter-particle collisions. For the decay rate
$\gamma = 1/\tau$ they found a linear dependence $\gamma = \gamma_0 + c\,
\density$ on the background gas density $\density$ which can be explained using
classical scattering theory. However, regarding the vibrational ground state in
the limit $\density \to 0$, the background atoms cannot play a role in the
reduction of the lifetimes. Nevertheless, the measurements indicate a ground
state's lifetime of $\tau_0 = 47.6\,\mu$s while the bare atoms have a lifetime
of $\tau_\text{atom} = 62.5\,\mu$s. In this zero density limit, the system only
consists of three parts, namely the Rydberg core, the Rydberg electron and the
\GSA, so, obviously, the reduced lifetime must be caused by the \GSA\ somehow
perturbing the Rydberg wave function.

In order to uncover the reason for this difference in the lifetimes of about
25\%, a convenient approach in molecular physics would be to evaluate the
non-adiabatic energy terms which govern the energy transfer between nuclear and
electronical motion. These terms couple a particular Rydberg state with the
continuum of dissociation of a lower-lying one and, thus, allow for a
predissociation process under the change of the electronic state. In addition to
spontaneous photon emission and blackbody induced radiation, such a process may
further shorten the lifetime of the molecule. However, a quantitative evaluation
of the corresponding decay rate requires the determination of the non-adiabatic
coupling terms which is a difficult task.

As an alternative, we therefore persue a different approach in this paper to
investigate the decay mechanism initiated by the \GSA\ in a Rydberg molecule.
That is based on a quantum-classical description of the Rydberg molecule
\cite{junginger2012c} which allows us to take into account the effect of the
\GSA\ onto the Rydberg atom beyond the usual treatment of a static disturber. In
previous work \cite{Granger2001, Li2011} the perturbation of the Rydberg state
has been considered at a fixed position $\RR$ of the non-moving \GSA. In this
paper, we introduce a model in which the moving of the \GSA\ plays an important
role, viz.\ the Rydberg state is continuously detuned due to a transfer of
kinetic energy from the \GSA\ to the Rydberg state. A detuning comparable or
larger than the natural linewidth of the Rydberg state will enhance the
transition to a lower Rydberg state resulting in the decay of the molecule by
either spontaneous photon emission or predissociation. As we will show below,
the time when the detuning significantly exceeds the natural linewidth of the
vibrational ground state agrees very well with the lifetime in the zero density
limit. Moreover, we find that the detuning becomes smaller with increasing
principal quantum number, so that the experimentally measured tendency of longer
lifetimes with increasing $n$ is also in accordance with our treatment.

Our paper is organised as follows: In Sec.\ \ref{sec-theory}, we give a brief
review of the quantum mechanical description \cite{Greene2000} as well as of the
quantum-classical treatment \cite{junginger2012c}. In Sec.\ \ref{sec-results},
we apply the latter approach to a rubidium Rydberg molecule in the s-state. The
detuning is discussed in detail for the vibrational ground state of a molecule
with $n=35$ and comparisons with the corresponding experiment are made.
Moreover, we demonstrate the effect of the detuning for different experimentally
accessible principal quantum numbers $n=33 - 41$.

\section{Theory}
\label{sec-theory}

In this section, we review the quantum mechanical description of the scattering
process between the Rydberg electron and the \GSA\ as well as the
quantum-classical treatment. Both approaches are discussed briefly and we refer
the reader to Refs.\ \cite{Greene2000,junginger2012c} for details.

\subsection{Molecular potential of Rydberg molecules}

\begin{figure}[t]
\centering
\includegraphics[width=.7\columnwidth]{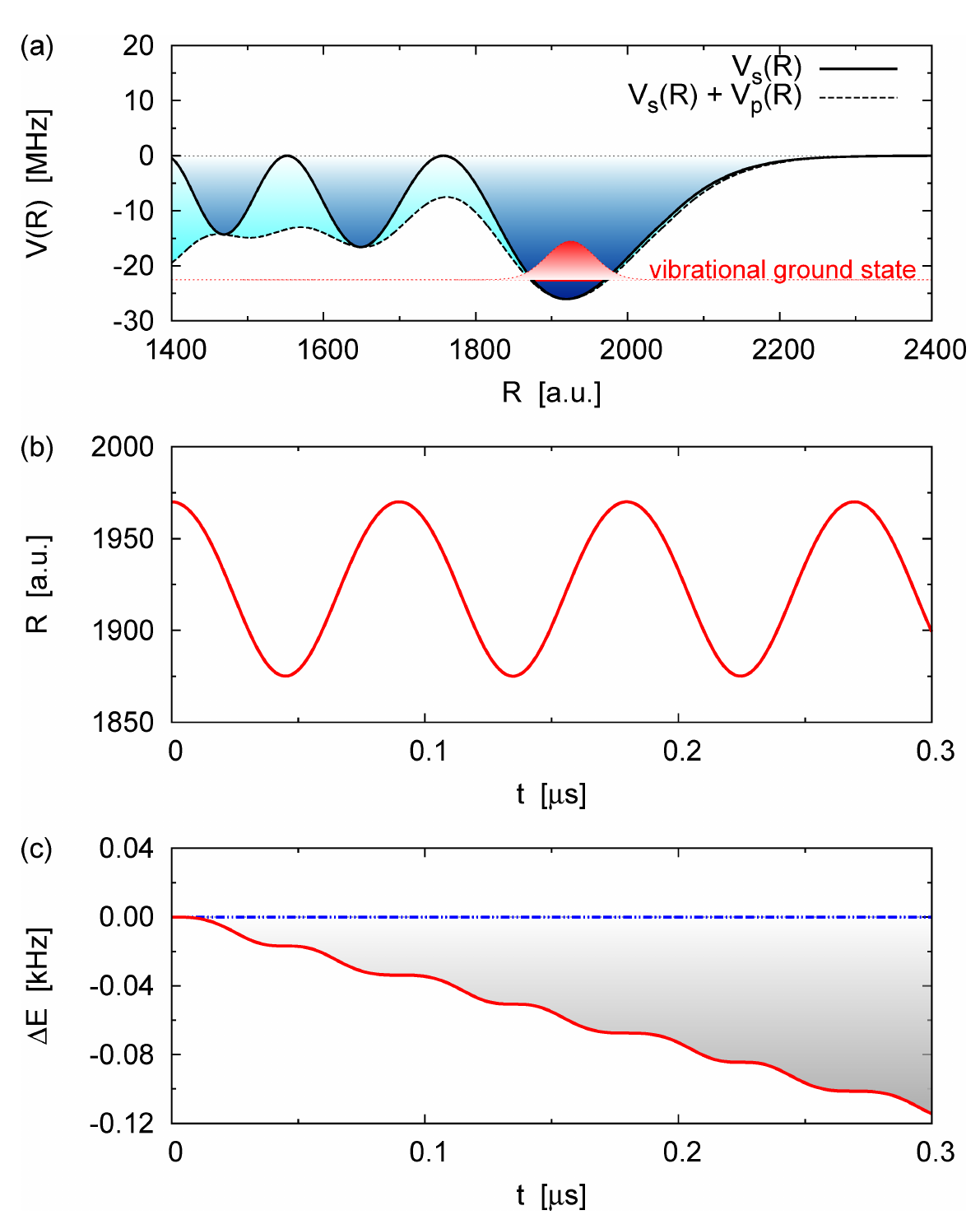}
\caption{
(a) Molecular potential of a rubidium Rydberg molecule in the $n=35$ s-state.
Shown are, both, the potential $V_\text{s}(\RR)$ taking into account only the
s-wave contribution (solid line) and for comparison the potential
$V_\text{s}(\RR)+V_\text{p}(\RR)$ including p-wave scattering (dashed line). At
internuclear separations $R\approx 1875-1970\,$a.u.\ where the vibrational
ground state (horizontal red line) is located the p-wave contribution can be
neglected. %
(b) Oscillation of a point particle in the molecular potential $V_\text{s}(\RR)$
at the energy of the vibrational ground state. %
(c) Energy detuning $\Delta E$ of the \GSA: Within the conservative potential
resulting from the mean-field approximation, there is no energy detuning (blue
dashed-dotted line), however, energy is transferred to the Rydberg electron with
each oscillation in the quantum-classical treatment (red solid line). Note that,
because of energy conservation, the energy of the Rydberg state is increased by
$-\Delta E$.
}
\label{fig-pot-trajec-energy}
\end{figure}

The binding mechanism of the Rydberg molecule is based on the polarisation of
the \GSA\ under the influence of the \RE. This interaction can be described
theoretically using a Fermi type pseudopotential \cite{Fermi1934} 
\begin{equation}
	V(\rr,\RR) = 2 \pi a_\text{s}(k) \, \delta(\rr-\RR),
	\label{eq-Fermi-pseudopotential}
\end{equation} %
where $\rr$ and $\RR$ denote the positions of the Rydberg electron and the \GSA,
respectively. The whole information on the scattering process is given, here, by
the s-wave scattering length $a_\text{s}(k)$. It depends on the wave vector $k$
of the Rydberg electron and can be expressed by its first-order approximation
\cite{Omont1977} $ a_\text{s}(k) = a_\text{s,0} + \frac{\pi}{3} \alpha k +
\order{k^2}$ with $a_\text{s,0}=-16.05\,$a.u.\ being the zero-energy scattering
length and $\alpha=319$ the polarizability of the rubidium target
\cite{Bendkowsky2009, Bendkowsky2010, Molof1974}. Greene \etal\
\cite{Greene2000} showed that, in a mean-field approximation, this contact
interaction leads to the molecular potential
\begin{equation}
	V_\text{s}(\RR) = 2 \pi a_\text{s}(k) \abs{\psiryd(\RR)}^2,
	\label{eq-V-s}
\end{equation} 
where $\psiryd (\RR)$ is the value of the Rydberg wave function at the position
$\RR$ of the \GSA. In the case of $a_\text{s}(k) < 0$, we obtain an attractive
interaction which allows for bound states. Fig.\ \ref{fig-pot-trajec-energy}a
shows the resulting molecular potential \eqref{eq-V-s} for the above mentioned
physical parameters (solid line) and indicates the energy and internuclear
distance of the vibrational ground state (horizontal red line) which is located
in the outermost potential minimum.

Note that significant modifications of this potential are caused for small
internuclear separations when also p-wave scattering is taken into account by
the additional term $ V_\text{p}(\RR) = 6 \pi a_\text{p}^3 \abs{\nabla \psi_{\rm
Ry}(\RR)}^2$, with the p-wave scattering length $a_\text{p} = -21.15\,$a.u.\
\cite{Bendkowsky2010} (see dashed line in Fig.\ \ref{fig-pot-trajec-energy}a).
This correction is important for the formation of Rydberg molecules at energies
$E \lesssim 0$ \cite{junginger2012c}. However, in this paper we will solely
consider the vibrational ground state of the molecule which is located in the
outermost potential minimum, and, as can be seen in Fig.\
\ref{fig-pot-trajec-energy}a, the p-wave contribution can be neglected there.

\subsection{Energy detuning of the Rydberg state in the quantum-classical
framework}

One consequence of the mean-field approach of Greene \etal\ \cite{Greene2000} is
the fact that Eq.\ (\ref{eq-V-s}) associates a \emph{fixed} position $\RR$ of
the \GSA\ with the potential energy $V_\text{s}(\RR)$. Physically this is
equivalent to the assumption of a vanishing ratio of masses of the two
scattering partners, i.e.\ $\me/m_\mathrm{Rb} = 0$. Although the ground state
atom is, by far, heavier than the electron ($\me/\mRb \approx 6\times 10^{-6}$
for rubidium), this assumption is not strictly fulfilled. Considering the single
scattering events leading to the bound molecule, we must, therefore, expect a
momentum and, with it, also an energy transfer between the \GSA\ and the Rydberg
electron occurring with each orbit of the latter.

In order to take into account such effects, we will investigate the system in a
quantum-classical way \cite{junginger2012c}. We describe the positions and the
motion of, both, the Rydberg electron as well as the \GSA\ as that of point
particles while the scattering process between these two is treated fully
quantum mechanically. Because of the high excitation of the Rydberg atom ($n\gg
1$), correspondence principle allows us to treat the motion of the Rydberg
electron in terms of the classical trajectories whose angular momentum $L$ and
energy $E $ will be quantised semiclassically:
\begin{equation}
	L = l+\frac{1}{2}, \quad
	E = \frac{\vec{p}^2}{2} - \frac{1}{r} = -\frac{1}{2n^2}.
\end{equation}
Here,  $n=1,2,3,\ldots$ and $l=0,1,2,\ldots$ denote the principal and azimuthal
quantum numbers, $\vec{p}$ is the momentum of the Rydberg electron and
$r=\abs{\rr}$ is its distance from the core. 

Since the interaction between the Rydberg electron and the \GSA\ is of
contact-like type, we only need to consider those orbits which include the
position $\RR$ of the \GSA. Considering a Rydberg atom in an s-state ($l = m =
0$), there remain four ellipses which we \emph{locally} approximate by a
superposition of plane waves
\begin{equation}
	\psiryd (\rr) \approx \sum_{i=1}^{4}	\psipw^{(i)} (\rr) =
        \sum_{i=1}^{4} A^{(i)} \exp\left( \ii \vec{p}^{(i)} \vec{r}
        \right).
\end{equation} 
In order to determine their amplitudes $A^{(i)}$, we require this superposition
to fulfil
\begin{subequations}%
\begin{align}
  \psipw |_{\rr=\RR} &= \psiryd |_{\rr=\RR}\;,
\label{eq-system-of-equations-a}\\
  \partial_{\rho,z} \psipw |_{\rr=\RR} &= \partial_{\rho,z} \psiryd
|_{\rr=\RR}\;,
\label{eq-system-of-equations-b}\\
  \partial^2_{\rho,z} \psipw |_{\rr=\RR} &\approx \partial^2_{\rho,z} \psiryd
|_{\rr=\RR}\;,
\label{eq-system-of-equations-c}	
\end{align}%
\label{eq-system-of-equations}%
\end{subequations}%
i.e.\ to reproduce the value of the Rydberg wave function and its derivatives
identically on the one hand, and to approximately reproduce the second
derivatives on the other. Here, Eq.\ \eqref{eq-system-of-equations-c} has to be
understood in the sense that the norm of the difference of both sides be
minimal. Altogether, Eqs.\
\eqref{eq-system-of-equations-a}--\eqref{eq-system-of-equations-c} provide the
best possible approximation of the Rydberg wave function by the four plane
waves.

With the knowledge of the plane waves, the scattering interactions can be
calculated straightforwardly: Each of them represents a Rydberg electron with
momentum $\vec{p}_\mathrm{in}^{(i)} = \me \ve^{(i)}$ which is scattered to an
outgoing wave
\begin{equation}
  \psi_\mathrm{out}^{(i)} \sim \frac{\exp\bigl( \ii\,
    {p}_\mathrm{out}^{(i)} \abs{\rr-\RR} \bigr)}{\abs{\rr-\RR} } .
  \label{eq-psi-out}
\end{equation}
Because of the spherically symmetric angle distribution the total momentum of
the outgoing s-wave is $\vec{p}_\mathrm{out}^{(i)}=0$, and the momentum transfer
$\Delta p$ is connected with an energy transfer $\Delta E$ between the \GSA\ and
the Rydberg electron.

Describing successive collisions of $N^{(i)}$ electrons by a current density
$\vec{j}^{(i)}= n_\mathrm{e}^{(i)} \vec{v}_\text{e}^{(i)}$
($n_\mathrm{e}^{(i)}=|A^{(i)}|^2$ is the electron density on the $i$-th Kepler
ellipse), the total momentum transfer $\Delta \vec{p}^{(i)} = N^{(i)} \me
\ve^{(i)} $ over time can be associated with a classical force
\begin{equation}
  \vec{F}^{(i)}
  = \Delta \vec{p}^{(i)}/{\Delta t}
  = n_\mathrm{e}^{(i)} \me \sigma
      \abs{\vec{v}_\text{e}^{(i)} - \vec{v}_\mathrm{Rb}}^2
      \hat{\vec{e}}_{(\vec{v}_\text{e}^{(i)} - \vec{v}_\mathrm{Rb})}
\label{eq-force}
\end{equation}%
acting on the \GSA. Here, $\sigma=4\pi a_\text{s}^2(k)$ is the total scattering
cross-section and $\hat{\vec{e}}_{(\ve^{(i)} - \vRb)}$ is the unit vector in the
direction of $\ve^{(i)} - \vRb$. The energy detuning over time of the \GSA\
induced by Eq.\ \eref{eq-force} is consequently given by
\begin{equation}
\frac{\dd E}{\dd t}
= \sum_{i=1}^4 \vec{F}^{(i)} \, \vRb
= \me \sigma \sum_{i=1}^4 n_\mathrm{e}^{(i)}
	\abs{\ve^{(i)} - \vRb}^2
	\left[ \hat{\vec{e}}_{(\ve^{(i)} - \vRb)}
	\, \vRb \right] \, ,
\label{eq-detuning}
\end{equation}%
where the sum $i=1,\ldots,4$ takes into account all four contributing Kepler
ellipses. Note that, because of energy conservation, the detuning of the Rydberg
state is given by the negative of Eq.\ \eref{eq-detuning}, and that the value of
$\dd E/\dd t$ is significantly changing with time because of the vibrational
motion of the \GSA. Thus, an appropriate mean energy detuning can be obtained by
averaging Eq.\ \eref{eq-detuning} over one oscillation,
\begin{equation}
\langle \dd E/\dd t \rangle = \int_0^T \dd t'\, [\dd E/\dd t'],
\label{eq-detuning-average}
\end{equation}
where $T$ is the periode of the vibrational oscillation. Moreover, since the
energy detuning is proportional to the velocity of the \GSA , $\dd E/\dd t
\propto \vRb$, the main contribution is obtained when the \GSA\ is fast and
there will be no contribution when it does not move, i.e.\ at the turning points
of the oscillation.

Depending on whether the two scattering partners fly in the same ($\ve^{(i)}
\cdot \vRb >0$) or in the opposite ($\ve^{(i)} \cdot \vRb <0$) direction, the
respective contributions to Eq.\ \eqref{eq-detuning} differ and the latter is
always dominant because in this case the term $|\ve^{(i)} - \vRb | $ is larger.
Therefore, the net effect of Eq.\ \eqref{eq-detuning} is always an energy
transfer from the ground state atom to the Rydberg electron, i.e.\ the Rydberg
state is permanently detuned.

The energy detuning is quantum mechanically not strictly forbidden because the
vibrational ground state of the Rydberg molecule is metastable and has some
finite lifetime $\tau$. This directly implies a finite width in frequency space
$\Delta \nu = (2 \pi \tau)^{-1}$ which is, for typical lifetimes of a few tens
of $\mu$s, on the order of several kHz, and, thus, small compared to the level
spacing of the Rydberg atom. Note that an exponential decrease of the ground
state's population leads to a Lorentzian form 
\begin{equation}
	I(\nu) \sim \frac{1}{(\nu - \nu_0)^2 + (\Delta \nu /2)^2}
	\label{eq-lorentzian}
\end{equation}
of the spectroscopic line. Consequently, detunings $\nu \neq \nu_0$ are allowed,
but reduce the intensity $I(\nu)$ and, for very low intentities, one expects a
change of the electronic state (e.g.\ $n \to n-1$). Because the intensity
$I(\nu)$ decreases continuously with increasing difference of the frequency $\nu
- \nu_0$ and the tail of the Lorentzian is extended infinitely, a clear
threshold for the frequency $\nu$ which distinguishes the quantum mechanically
allowed region $\nu$ (where we do not expect a change of the electronic state)
from the forbidden one (where we do expect the electronic state to change) does,
therefore, not exist. Therefore, we estimate these regimes in the following way:
A detuning which is smaller than the natural linewidth, $|\nu - \nu_0| \lesssim
\Delta \nu$, will be surely allowed, while much larger detunings $|\nu - \nu_0|
\gg \Delta \nu$ are quantum mechanically forbidden so that we expect a detuning
on the order of a few natural linewidths, $|\nu - \nu_0| \sim \Delta \nu$, to
determine the threshold where the transition of the electronic state will set in
resulting in the decay of the molecule by spontaneous photon emission or
predissociation.

\section{Results and discussion}
\label{sec-results}

In this section, we demonstrate the effect of the energy detuning in Eq.\
\eqref{eq-detuning} for $^{87}$Rb Rydberg atoms in the spherically symmetric
s-state. At first, we will discuss the detuning of the vibrational ground state
in general, then compare the results with the experimentally measured lifetimes
and, finally, investigate the effect of different principal quantum numbers $n$
on the time development of the detuning. Note that all calculations presented
take into account a quantum defect correction of $\delta=3$ \cite{Li2003}.

\subsection{Detuning of the molecular states}

The Rydberg molecule is quantum mechanically described by the corresponding wave
function $\psiryd$. The vibrational ground state of the molecule is located in
the outermost potential minimum of the oscillatory molecular potential and does
not extend into one of the neighbouring wells, since their minima are
energetically higher than the ground state (see Fig.\
\ref{fig-pot-trajec-energy}a). This wave function describes the probability
density of the \GSA, which oscillates in this outermost potential well, to be
found at a specific internuclear distance.

The classical analogue is a point particle oscillating in this potential well
between two turning points $R_\text{min}$ and $R_\text{max}$. Fig.\
\ref{fig-pot-trajec-energy}b shows this oscillation of the ground state atom
with a binding energy of $E_0 = -22.5174\,$MHz \cite{Jonathanprivate} for a
Rydberg molecule in which the Rydberg atom is excited to the $n=35$ s-state. In
this case, the turning points are $R_\text{min} \approx 1875\,$a.u.\ and
$R_\text{max} \approx 1970\,$a.u.

In the framework of the conservative mean-field potential \eqref{eq-V-s} this
oscillation would continue until $t \to \infty$ and the energy of the ground
state atom would not change, i.e.\ $\Delta E = 0$ (see blue dashed-dotted line
in Fig.\ \ref{fig-pot-trajec-energy}c). However, this situation is different in
the quantum-classical treament (see red solid line in Fig.\
\ref{fig-pot-trajec-energy}c): With each oscillation, i.e.\ when $\vRb \neq 0$,
energy is transferred to the Rydberg electron, and the \GSA\ loses energy. The
total detuning is on the order of $\sim 0.02\,$kHz per oscillation and, thus,
very small compared to the binding energy of several MHz. The detuning per
oscillation therefore remains almost constant with time, and from averaging Eq.\
\eref{eq-detuning} we obtain a mean detuning over time [see Eq.\
\eqref{eq-detuning-average}] on the order of $\langle \dd E/ \dd t \rangle \sim
0.3\,$kHz/$\mu$s.

\subsection{Comparison with the experiment}

In this section, we compare the time development of the detuning with the
experimental investigations of the lifetimes. We again consider a Rydberg
excitation of $n=35$ which has also been investigated in detailed experimental
lifetime measurements by Butscher \etal\ \cite{Butscher2011}. 

\begin{figure}[t]
\centering
\includegraphics[width=.7\columnwidth]{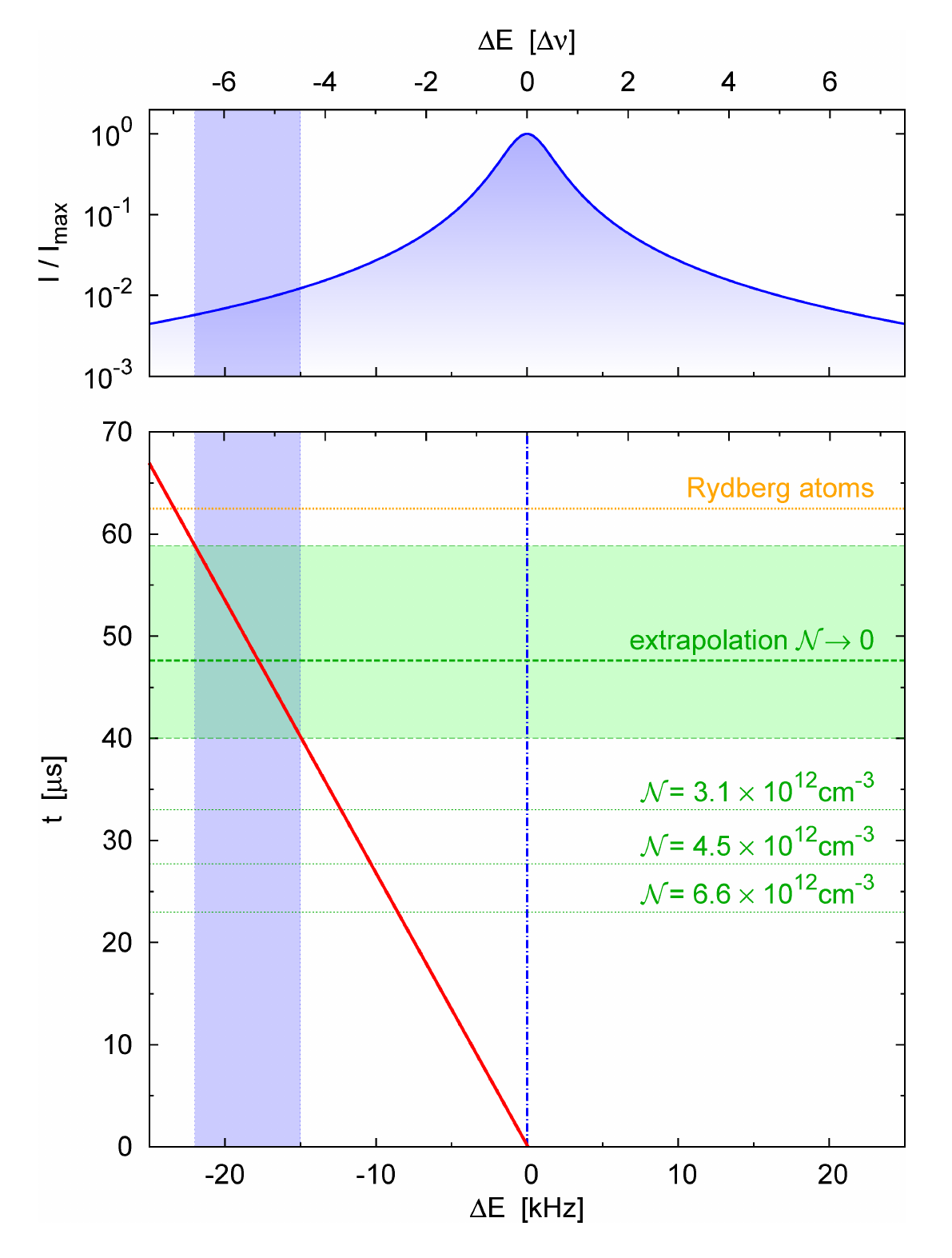}
\caption{
(Top) Lorentzian peak with a width $\Delta \nu$ corresponding to the lifetime of
$^{87}$Rb Rydberg molecules in the zero density limit. 
(Bottom)
Development of the \GSA's energy, $\Delta E$ (again, the Rydberg state is
detuned by $-\Delta E$): Within the conservative mean-field description of the
potential $V_s(\RR)$ in Eq.\ \eqref{eq-V-s} the energy remains constant (blue
dashed-dotted line), i.e.\ $\Delta E = 0$. By contrast, the quantum-classical
model shows a permament detuning (red solid line). For comparison the
experimentally measured lifetimes of the Rydberg atoms and molecules are shown
(horizontal dotted lines) for different densities $\density$ as well as in the
limit $\density \to 0$. The shaded areas indicate the experimental error (green
horizontal stripe) and the region of the energy detuning which is present in
this regime (blue vertical stripe). See text for further description.}
\label{fig-extrapolation}
\end{figure}

Fig.\ \ref{fig-extrapolation} again shows the time development of the detuning
(red solid line). The authors of Ref.\ \cite{Butscher2011} have measured the
lifetimes of the $n=35$ Rydberg state for different densities $\density$ of the
atomic background gas and found that the lifetime decreases with increasing
density $\density$. Some of their results have been included in Fig.\
\ref{fig-extrapolation} for comparison (horizontal green dotted lines). As
discussed in the introduction, they found a linear dependence of the decay rate
on the density that indicates a $\tau_0 = 47.6\,\mu$s lifetime of the Rydberg
molecule in the zero density limit (horizontal green dashed line, the horizontal
green stripe indicates the experimental error). In contrast to the expectation
that, in this limit, one would obtain the lifetime $\tau_\text{atom} =
62.5\,\mu$s of the bare atomic state (horizontal orange dotted line), the
extra\-polated value is still significantly reduced by about 25\%.

To determine whether or not the detuning of the Rydberg state discussed above
can play a role in reducing the lifetimes, we compare these two on the relevant
time scale. As can be seen in Fig.\ \ref{fig-extrapolation} the detuning of the
Rydberg state reaches a value of $\Delta E \approx 15 - 22\,$kHz (indicated by
the vertical blue stripe) within the extrapolated lifetime including the
experimental error. Comparing this detuning with the corresponding Lorentzian
profile of the vibrational ground state which has a linewidth of $ \Delta \nu =
3.4\,$kHz (see top of Fig.\ \ref{fig-extrapolation}), we obtain that this
corresponds to a detuning of $\Delta E \approx ( 4.5 - 6.5) \Delta \nu$. This
value is in accordance with the above estimated detuning where the quantum
mechanically ``forbidden region'' is reached, since for such a detuning the
Lorentzian curve has decreased to $I / I_\text{max} \lesssim 0.01$. The Rydberg
wave function is then strongly detuned which can result in a significantly
higher probability for its decay than one would expect from spontaneous decay or
blackbody induced radiation. Therefore, we expect the lifetime of the Rydberg
molecule to be determined by the time necessary for the detuning to
significantly exceed the natural linewidth of the molecule.

As already discussed above, we also expect this strongly detuned Rydberg wave
function to cause a change in the electronic structure (e.g.\ $n \to n-1$)
which, depending on the final state, may allow for different reactions, e.g.\
the dissociation of the Rydberg molecule. This picture is supported by the
quantum mechanical interpretation of the process: In the case of a fixed \GSA\
\cite{Granger2001, Li2011}, the scattering of a Rydberg electron will, in
general, change its direction, but not its energy. This means that the
scattering process does not conserve the angular momentum but the principal
quantum number ($n=\text{const}$). However, in the case of a \GSA\ with
nonvanishing velocity ($\vRb \neq 0$), each scattering process causes a change
of the Rydberg electron's energy. For detunings which are small compared to the
natural linewidth, this is possible within the linewidth of the respective
Rydberg state $n$. Significant detunings on the order of the natural linewidth
or even larger, however, are quantum mechanically related to couplings to other
quantum states, i.e.\ the strongly detuned Rydberg state must, in general, be
coupled to states with other principal quantum numbers $n$. The coupling can
induce transitions to lower electronic states resulting in the decay of the
molecule, as already discussed above.

\subsection{Dependence on the principal quantum number}

Detailed investigations of the lifetimes of Rydberg molecules can only be found
for the $n=35$ s-state in the literature so far \cite{Butscher2011}. However,
already in the publication on their first experimental realisation
\cite{Bendkowsky2009}, a dependence of the lifetime on the principal quantum
number $n$ has been discussed. Although these experiments have been performed at
finite densities $\density$ and an extrapolation to zero density is not possible
due to the lack of data, the measured tendency of longer lifetimes with
increasing quantum number $n$ can also be expected to hold in the limit
$\density \to 0$. We, therefore, present the behaviour of the Rydberg electron's
detuning for different $n$ in the following.

In order to determine the detuning for different quantum numbers $n$, we proceed
as done for the $n=35$ state: We determine the molecular potential for the
respective quantum number and place the point particle representing the \GSA\ in
the outermost well with an energy corresponding to the respective vibrational
ground state. The average energy detuning $\langle \dd E / \dd t \rangle$ is
calculated using Eq.\ \eqref{eq-detuning-average}.

\begin{figure}[t]
\centering
\includegraphics[width=.7\columnwidth]{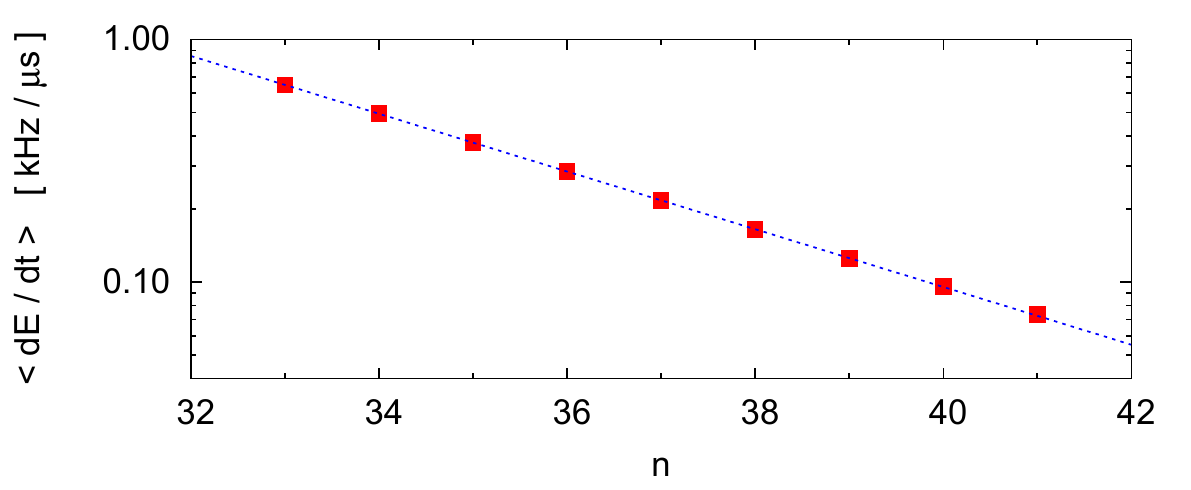}
\caption{
Dependence of the mean energy detuning over time $ \langle \dd E / \dd t
\rangle$ in Eq.\ \eqref{eq-detuning-average} on the principal quantum number
$n$. The calculations reveal a perfect exponential decrease of the detuning with
the quantum number $n$.
}
\label{fig-detuning}
\end{figure}

Fig.\ \ref{fig-detuning} shows this mean energy detuning over time for different
experimentally accessible quantum numbers $n = 33 - 41$ (red squares). The
figure illustrates that the value $\langle \dd E / \dd t \rangle$ strongly
depends on $n$ showing a significant decrease of the detuning with an increasing
quantum number. Two points contribute to this behaviour: (i) The extension of
the Rydberg electron's wave function scales with $\left< r \right> \sim n^2$ so
that the atom becomes more extended with increasing $n$. At the same time the
larger extension results in a smaller electron density $n_e^{(i)}$ at the
position of the \GSA, which lowers the effect of the finite mass correction term
in Eq.\ \eqref{eq-detuning}. (ii) For higher principal quantum numbers, the
outermost potential minimum becomes less deep so that the velocity of the \GSA\
decreases. Because $\vRb$ directly enters the correction term, its effect is
further reduced. 

For the decrease of the detuning with increasing $n$, we obtain a value of
\begin{equation}
(\Delta E / \Delta t)_{n} \sim 0.76^n
\end{equation}
from Fig.\ \ref{fig-detuning}. Assuming that -- in analogy to the calculations
for the $n=35$ s-state -- the decay of the Rydberg molecule was generally
induced by the \GSA\ when a detuning of $\Delta E \approx 15-22\,$kHz is
reached, this would mean an increase of the Rydberg molecule's lifetime by a
factor of $1/0.76 \approx 1.32$. Note that, considering the experimental error
stripes presented in Ref.\ \cite{Bendkowsky2009}, this value is in accordance
with the lifetime measurements of the Rydberg molecules by Bendkowsky \etal\ at
finite densities \cite{Bendkowsky2009}.

\section{Conclusion and outlook}

We have investigated the vibrational ground state of Rydberg molecules within a
quantum-classical treatment. In this model the dynamics of the \GSA\ and the
Rydberg electron are coupled, leading to a continuous energy transfer from the
\GSA\ to the Rydberg electron by which the Rydberg state is permanently detuned.

Comparing the time evolution of the detuning with these reduced lifetimes of the
molecule, we observe that the measurements agree very well  with the time at
which the detuning significantly exceeds the natural linewidth. We, therefore,
expect this detuning of the Rydberg state to give an important contribution to
the reduced lifetimes of Rydberg molecules. Also the experimentally measured
tendency of longer lifetimes with higher Rydberg excitation could be verified
within this model. However, further detailed experimental investigations of the
corresponding lifetimes need to be performed for different densities of the
background atomic gas as well as different quantum numbers $n$ to verify the
results also in the zero density limit of the atomic background gas. A fully
quantum mechanical treatment of the process discussed in this paper and the
accurate computation of the lifetimes are a challenge for future work.

\section*{Acknowledgements}
This work was supported by Deutsche Forschungsgemeinschaft. A.\,J.\ is grateful
for support from the Landesgraduiertenf\"orderung of the Land
Baden-W\"urttemberg. We also thank T.\ Pfau and his group for fruitful
discussions and support.

% literature ======================================
\section*{References}
% \bibliographystyle{unsrt}
% \bibliography{Literature.bib}

\end{document}